\documentclass[epj]{svjour}
\usepackage{amsmath}
\usepackage{amssymb}
\usepackage{graphicx}
\usepackage{color}
\usepackage{url}

\begin{document}

\title{Space use by foragers consuming renewable resources}
\author{
	Guillermo Abramson\inst{1}\thanks{\email{abramson@cab.cnea.gov.ar}}
\and
    Marcelo N Kuperman\inst{1}\thanks{\email{kuperman@cab.cnea.gov.ar}}
\and
    Juan M Morales\inst{2}\thanks{\email{jm.morales@conicet.gov.ar}}
\and
	Joel C Miller\inst{3}\thanks{\email{joel.c.miller@gmail.com}}
}
\institute{
Centro At\'omico Bariloche, CONICET and Instituto Balseiro, S. C. de Bariloche, Argentina
\and
Laboratorio Ecotono, INIBIOMA, CONICET and Universidad Nacional del Comahue, S. C. de Bariloche, Argentina
\and
Department of Biology, Pennsylvania State University, Pennsylvania, USA
}%

\date{Received: date / Revised version: date}

\abstract{
We study a simple model of a forager as a walk that modifies a relaxing  substrate. Within it simplicity, this provides an insight on a number of relevant and non-intuitive facts. Even without memory of the good places to feed and no explicit cost of moving, we observe the emergence of a finite home range. We characterize the walks and the use of resources in several statistical ways, involving the behavior of the average used fraction of the system, the length of the cycles followed by the walkers, and the frequency of visits to plants. Preliminary results on population effects are explored by means of a system of two non directly interacting animals. Properties of the overlap of home ranges show the existence of a set of parameters that provides the best utilization of the shared resource.}

\maketitle
\section{Introduction}

Animals usually exhibit complex patterns of movement which arise arise from the interaction between the individual and the environment \cite{turc98}. The motivations for movement depend on the animal's internal state (satiation, reserves, etc.), on the interactions with members of their own or other species and on previous experiences \cite{nath08,mor10}. Importantly, the way animals move affects how individuals redistribute themselves over space and thus has the potential to affect many ecological processes \cite{kareiva95,mor10}.

A broad group of animals move around in order to collect food from patches of renewable resources such as fruits, nectar, pollen, seeds, etc. For these animals we expect that their movement trajectories will depend strongly on the spatial arrangement of such patches \cite{cresswell97,ohashi07}. Often these animals play an important ecological role as part of mutualistic interactions, as they pollinate or disperse the seeds of the plants they visit. For seed dispersal in particular, empirical and theoretical studies show that the spatial distribution of plants contributes to the seed deposition patterns through its effect on animal movement \cite{leve05,kwit04,rus04,sara04,mor06,rus06,car08,lenz10,her11}. Understanding the emergence of space use of animals foraging for renewable resources, besides being an interesting theoretical topic, can allow us to build better studies of animal-plant interactions.

Previous studies considering systems of animals foraging on renewable resources have focused on finding optimal search strategies under different assumptions of animal perception and memory \cite{bartumeus02,barton09,fronhofer13}. It is clear that some animals are capable of finding profitable routes without a lot of computational power \cite{zollner99}. Also, much discussion has been devoted to animals' search paths and whether L\'evy walks or flights are predominant in nature \cite{viswanathan96,benhamou07,edwards07,reynolds12,boyer09}.

However, less attention has been paid to the emergence of space use patterns as the result of an interaction between the behavioral rules used by an organism and the spatial structure of the environment. In this work we use a simple model of foraging animals while traversing a territory populated by their source of sustenance. It is effectively a walker on a dissordered substrate, which is an interesting mathematical problem even in simple formulations, whose statistical properties has only recently been subject of investigation \cite{freund92,lima01,risau03}. Our main goal is to understand the emergent properties of simple movement rules of animals foraging for patchy and renewable resources.


\section{Model definition and dynamics}
\subsection{The substrate}

We consider a finite spatial domain, within which there are $N$ patches of vegetation, that the animal can visit to get food. These patches, which we may refer to informally as ``plants'' below, do not overlap and each one is endowed with a load of fruit $f_i(t)$, which is the only resource for the animals to consume in the system. Initially the plants have $f_i(0)=k_i\in (0,1)$, with uniform distribution.

We assume that the time scale of the vegetation dynamics is much slower than that of animal foraging and resource renewal. Thus, the number and position of patches remain constant. We have used two distinct distributions:

\begin{itemize}
\item The space is the unit square, and the plants are set at uniformly distributed points on the square.

\item The space is of undefined size, on which a first plant is set at random. The rest of the plants are set sequentially, at a distance from the previous one which is randomly drawn from a lognormal distribution (and uniformly distributed in azimuth).
\end{itemize}

The lognormal distribution produces hierarchically clustered substrates, compared to the Poissonian distance between plants resulting from the uniform one. For the sake of brevity we will show mainly the results corresponding to uniform substrates, with a few mentions to the differences arising in the lognormal ones.

The only dynamics of the vegetation substrate is a continuous replenishment of the fruitload of each plant, according to an exponential growth that saturates to the initial value $k_i$. This simple relaxation dynamics can represent a ripening process, for example, in such a way that the fruit load available to the animals is only the ripe fruit. Since we are, at the present stage, only interested in the short time scale of the dynamics, we consider no seasonality nor density dependence in this dynamics. If the time of the most recent visit of an animal to plant $i$ is $t_v$, the fruit at any time $t>t_v$ (before a new visit) grows as:
\begin{equation}
f_i(t) = k_i + (f_i(t_v)-k_i)\,e^{-(t-t_v)/\tau}.
\end{equation}
All plants relax with the same ripening time $\tau$, which is one of the major control parameters of the system.

\subsection{The walker}

An animal behaves as a foraging walker on the vegetation substrate which, being closed and finite, effectively works as an enclosure. It browses from plant to plant eating fruit. This dynamics is kept as simple as possible, without any transit time between plants nor perching time while feeding. When the animal gets to a plant it reduces the fruitload an amount of $b$. Immediately after eating it chooses the next plant and goes there. Furthermore, we consider no satiation and no rest. The next plant is chosen according to a stochastic rule of proximity and fruitload, as described below.

Our main assumption about the movement is that the main factor of the movement is proximity of the food. This is in fact the case in many foraging species, in particular when the distribution of the food resource is not heterogeneous in the extreme. It is, however, unreasonable to suppose that the animal will visit any site only because it is nearest, disregarding its food availability. We assume, then, that it will take that step only if it perceives that the plant has enough food. Also, we assume that the walker does not exhaust the available resource at a site. This is also reasonable because animals have a limited gut capacity, they might not want to stay at a site too long for fear of predation, etc. Rather, after taking a fraction of the available fruitload, it moves on to the next chosen site \cite{mor06,car08}.

These ingredients take the following mathematical form in the model. The animal checks the distances from the plant it's occupying, and chooses where to go next with an exponentially decaying probability depending on their order of distance. If the probability to visit the nearest plant is $P(1)=r$, then the animal chooses the $n$-th plant in order of distance with probability $P(n)=r(1-r)^{n-1}$. Observe that if $r=1$ the probability of visiting the nearest plant is 1, and the rule becomes deterministic. In this case, the animal always chooses the nearest plant. If $r<1$ there is a finite probability of visiting \emph{any} plant in the system, with closer ones being favorites. Note, also, that the probability does not depend on the \emph{distance}, but on the \emph{order} of distances. The rationale behind this is that the animal \emph{needs} to go somewhere to get its food, disregarding the distance. So, it doesn't matter if the $n$-th patch is close or far away: the animal will choose it with the same probability. A probability distribution that \emph{depends} on the distance (associated with a cost of reaching it, for example) could give a different phenomenology. This will be explored elsewhere.

After the animal chooses a plant according to the geometric distribution just explained, it checks whether this plant has enough food. If the current fruitload of the plant exceeds a threshold $u$, the animal moves to it. If the fruitload is less than $u$ it regards the plant as worthless, and chooses the next in the order of distance, checking again whether this next plant has enough fruit.

In summary, the random walk rules are:
\begin{enumerate}
\item Choose the next plant according to the geometric distribution of their order of distance.
\item \label{check} Check the fruitload of the chosen plant and:
\begin{enumerate}
\item If the load is greater than $u$, keep this choice.
\item If the load is smaller than $u$, choose the next plant in order of distance. Go to \ref{check}.
\end{enumerate}
\item Go to the chosen plant and reduce its load by $b$.
\end{enumerate}

The parameters $b$ and $u$ affect weakly the properties of the walk (results not shown), and have been kept constant in our simulations. Considering that the maximum fruitload is $f=1$, which can be interpreted as a cluster of fruit, we have used $b=0.1$ and $u=0.2$ throughout. The parameter $r$ affects the walk in a stronger way, and we have used it as the second (besides the relaxation time $\tau$) relevant control parameter of our analysis.

\section{A single animal}

\begin{figure}[t]
\begin{center}
\includegraphics[width=\columnwidth]{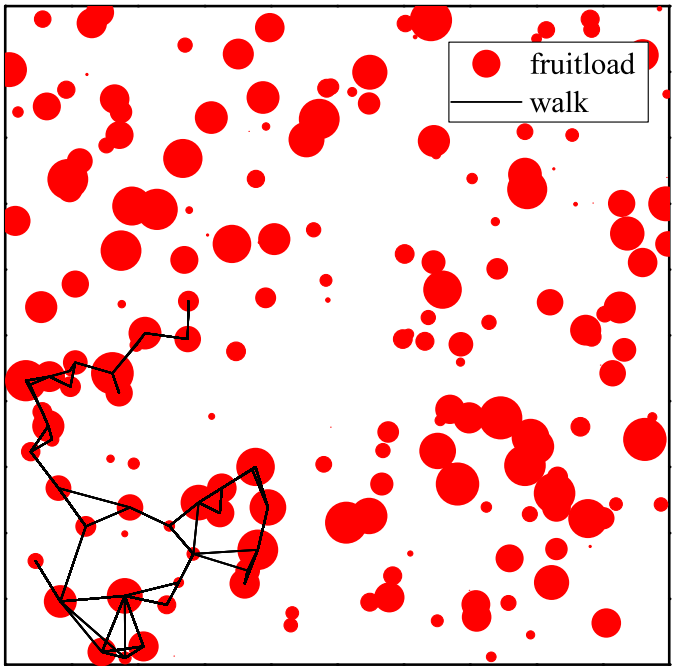}
\end{center}
\caption{Stationary track (i. e., discarding the transient) of a single animal on a uniform substrate. The stationary pattern is a cycle with period 553 steps.}
\label{tf500-1}
\end{figure}

Let us consider first a single animal in the system. In the  deterministic version of the rule, we observe that a periodic cycle arises (an example is shown in Fig.~\ref{tf500-1}, where the transient part of the walk has been discarded). Bear in mind that the walker has no memory, at variance with related models whith periodic behavior, such as the Tourist Walker of Ref.~\cite{lima01}. If stochasticity is allowed in the rule (so the animal can explore farther plants) the strict periodicity is broken, but some aspects of it persist, as will be discussed below. Let us exemplify these behaviors in different situations.

\begin{figure}[t]
\begin{center}
\includegraphics[width=\columnwidth]{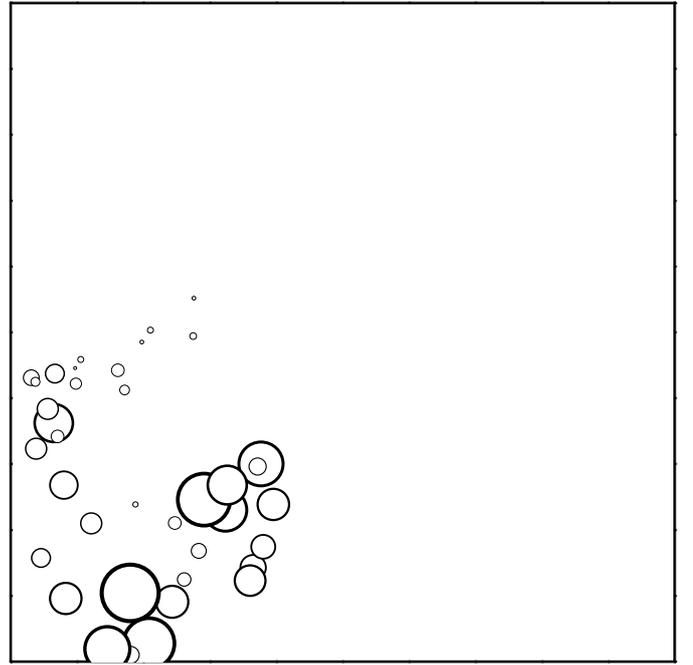}
\end{center}
\caption{Probability of occupation of space. The footprint of the home range occupies 12\% of the habitat plants. Circle size is proportional to the frequency of visits.}
\label{tf500-2}
\end{figure}

Consider the uniform random substrate shown in Fig.~\ref{tf500-1}. The size of the circles corresponds to the fruitload of each plant (in the relaxed state). A deterministic walker ($r=1$), after a short transient sets on a complex but cyclic track with a well defined home range. The black lines in the same panel show the steps of the walk, which has a period of 553 steps (of course, much less than 553 plants are visited during a cycle). Additionally, Fig.~\ref{tf500-2} shows the plants visited by the walker, indicating with the size of the circles the frequency of visit. It is clear that the use of the resource is very heterogeneous, both in space and in time. As this example shows, the frequency of visits is not simply correlated with the fruiload. It arises not only from the fruitload but also from the spatial context of the visited patch.

Having a single animal in the system provides a probe to reveal topological properties of the substrate. It is remarkable that the simple rules of this model, which do not consider any memory nor explicit cost in moving, are enough to guarantee the emergence of a finite home range.

Since the memory of the walk is effectively stored in the landscape, the relaxation time $\tau$ affects the length of the cycles and the fraction of used space. A longer ripening time makes fruit more difficult to find, and the home range expands considerably with correspondingly longer excursions. The track is still cyclic and periodic but the home range is traced in a very complex manner. A faster relaxation, in turn, has the effect of shrinking the home range and reducing the period of the cycle. In the case of lognormal substrates, the clustered organization of patches has the effect of confining the animal's home rage. In such cases most of the space use is restricted to one subcluster.

\subsection{Properties of periodic walks}

Even when the attractor for a particular set of parameters is periodic there is a great deal of variability in the duration of the period due to the strong dependence on the random substrate. Since both the period and the fraction of plants visited by the animal are measures of the use of the system we analyze here both magnitudes in a series of simulation runs. After discarding the transient we measure the period and the fraction of visited plants. The procedure was repeated for 1000 realizations, each one with a different substrate of $N$ plants and a different initial condition, and for a range of ripening times.

\begin{figure}[t]
\begin{center}
\includegraphics[width=\columnwidth]{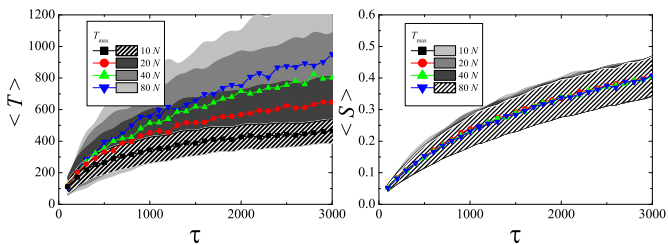}
\end{center}
\caption{Cycle length and fraction of visited plants as a function of ripening time. $N=200$, maximum observing times shown in the legend as multiples of $N$. Averages and standard deviations of 1000 realizations. Longest period detected: $1/2\,T_{max}$.}
\label{ciclo-tfrut}
\end{figure}

Figure \ref{ciclo-tfrut} (left) shows the average length of the cycle, $\langle T\rangle$, as a function of the ripening time $\tau$. After the transient, the walk was tested for periodic behavior of period $T$. With our method, the longest observable period can be detected as a single repetition of a pattern. As a consequence, the detection of longer periods can be affected by the maximum running time. Within this limitation, we have adjusted the observed time to avoid such an artifact as best as possible.

As anticipated above, for a given value of $\tau$ there is great variability in the distribution of periods. The shaded areas shown above and below the mean values should be interpreted as an inherent property of the distributions of cycle lengths. The size of these fluctuations depends on the system size and, consequently, may play a noticeable role in small systems. Since these distributions are broad, and our measurements correspond to a finite observation time, the averages are biased towards smaller values. For this reason we probed the paths with progressively longer observation times (using lengths of $aN$ with $a=10$, 20, 40 and 80, as indicated in the figure). These explorations show that there exist longer and longer periods.

The average length of the cycles grows slowly with $\tau$, and also with the size of the system (not shown). In some sense, the length of the cycle is combinatorial and it is not surprising that, as more space or time is available, longer cycles are possible. Interestingly, though, the number of visited plants does not share this property. Figure~\ref{ciclo-tfrut} (right) shows this \emph{home range size} $S$ as a function of $\tau$ for the same set of runs shown in the left panel. We see that the fraction of space used by the animal grows with $\tau$ as does the average cycle length. But it saturates for longer observation times (it's already asymptotic after the shortest observation of $10N$ time steps). We believe that this is an important result of our findings. Its robustness with respect to many details of a wide family of models suggests that  it may lie at the core of the origin of the existence of home ranges in animals that forage on renewable resources that are relatively fixed in space (i. e. with a slow resource dynamics).

The periodicity of the deterministic version of the walk can be understood as deriving from a finite set of available states in a closely related model, as follows. Assume that $b=u$ for simplicity, and that all plants are at their asymptotic levels $k_i$ at $t=0$. Let $p(t)$ denote the position of the walker at (discrete) time $t$, and $\chi_i \in \{1,0\}$ be the presence (1) or absence (0) at plant $i$, when the walker is at $p$. Then the amount of fruit at plant $i$ depends on previous visits and can be written as:
\begin{equation}
f_i(t)=k_i - b\sum_{t'=0}^{t-1} \chi_i(p(t'))\,e^{-(t-t')/\tau}.
\end{equation}
Because of the exponential relaxation the main contribution in this sum corresponds to recent times. If $t-t'$ is large, $b\,e^{-(t-t')/\tau}$ is small. In particular, for a given $\epsilon>0$, choose $T$ such that $k_i e^{-T/\tau}<\epsilon$ (for all $i$). This puts a bound on the contribution of the history previous than $T$ steps into the past:
\begin{equation}
b\sum_{t'=0}^{t-T} e^{-(t-t')/\tau} = b\,e^{-T/\tau}\,\sum_{t'=0}^{N} e^{-(t-T-t')/\tau},
\end{equation}
which can be made arbitrarily small by the choice of $\epsilon$. Since what has be substracted from plant $i$ at time $t-T$ is at most $k_i$ (its asymptotic value), the contribution of the history previous to $t-T$ is at time $t$ at most $\epsilon$:
\begin{equation}
f_i(t)-\epsilon < k_i - b\sum_{t'=t-N}^{t-1} \chi_i(p(t'))\,e^{-(t-t')/\tau} \le f_i(t).
\end{equation}
Then we can approximate:
\begin{equation}
f_i(t) \approx k_i - b\sum_{t'=t-T}^{t-1} \chi_i(p(t'))\,e^{-(t-t')/\tau},
\end{equation}
which tells that all that matters is the history going back $T$ units of time. With $N$ plants, there are $N^T$ possible histories, so the configuration space is finite, and periodicity follows.

As can be seen, for this proof to be rigorous it is necessary that $\epsilon$ is small enough that there is no difference in what choice the walker makes at each step. Given that the choice is made based on the rank of distances, this can always be ensured. If the choice were, instead, a continuous probability based on the distance, the argument would be weaker. (Still, it would be approximate and with a valid regime of applicability.) It is also apparent that the shape of the relaxation plays a role in this phenomenon: on the one hand, a longer $\tau$ increases the periods---as was already observed in the simulations above. On the other hand, a relaxation with a functional form slower than exponential---algebraic, for example---may hamper the existence of periodic trajectories.

\subsection{Properties of stochastic walks}

\begin{figure}[t]
\begin{center}
\includegraphics[width=\columnwidth]{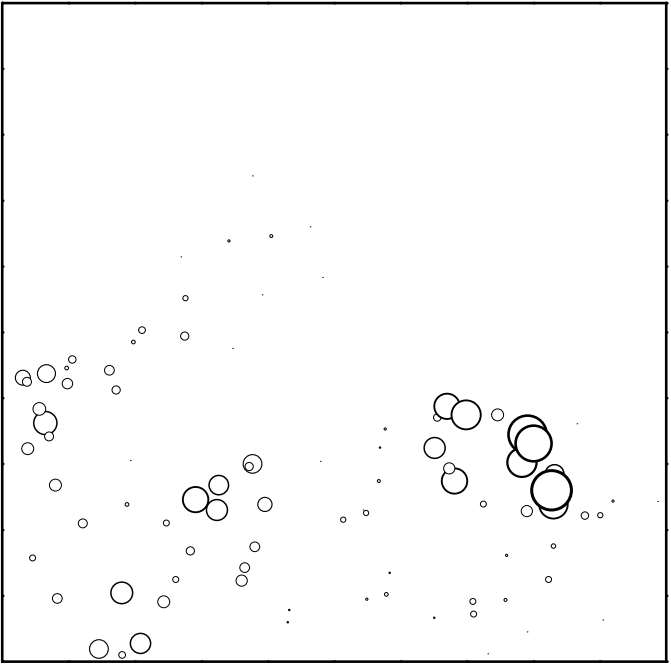}
\end{center}
\caption{Probability of occupation of space with $r=0.8$. The home range occupies now 43\% of the plants. Same substrate and other parameters as in Figs.~\ref{tf500-1} and \ref{tf500-2}.}
\label{aperiodic}
\end{figure}

A little randomness can help the animal browse its range more thoroughly, as shown in the example of Fig. \ref{aperiodic}. Here the same landscape and initial condition were used as in Figs.~\ref{tf500-1} and \ref{tf500-2}, but reducing the probability of choosing the nearest plant to $r=0.8$. The path is now \emph{non periodic}. Nevertheless, the space usage is still contained within a home range. The fraction of used plants is now 43\% of the total. The details are very dependent on the fluctuations of the substrate (which, with $N=200$ plants, are rather strong), but statistical conclusions can be derived nevertheless.

An important difference between these noisy walkers and the deterministic ones shown in the previous section is that the home range is not as well defined. The reason is that there is a small but finite probability of migrating to \emph{any} plant in the system. As a consequence of this, there is a slow drift away from the main track that may eventually cover the whole system. Preliminary observations show that the tracks behave as quasi-cycles that slowly drift in the substrate, but the whole phenomenology of this has not been completely explored at the present stage, and will be addressed elsewhere.

\begin{figure}[t]
\begin{center}
\includegraphics[width=\columnwidth]{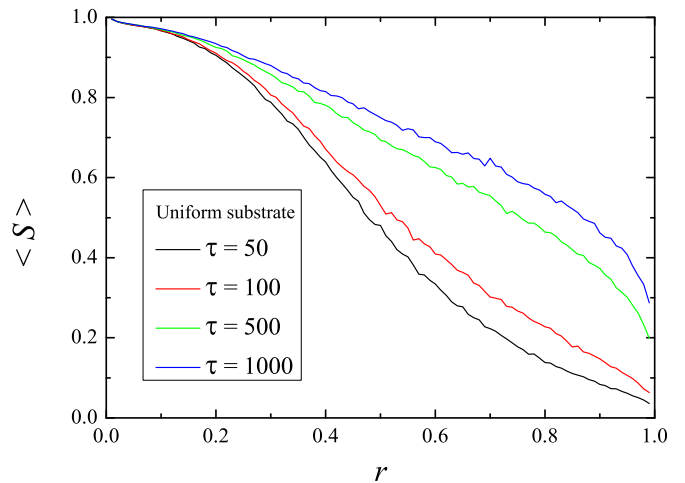}
\end{center}
\caption{Average home range size (measured as fraction of plants used) as a function of the noise parameter $r$ ($r=1$ corresponds to deterministic walks, with the walker always choosing the nearest plant; when $r=0$ it goes to any plant at random with uniform probability).  Substrates of 400 plants, and runs of $10N$ time steps, measured after a transient of $10N/4$ time steps. Each value of $r$ is the average of 1000 realizations (each with a different substrate).}
\label{uso-tfrut}
\end{figure}

The results presented in Fig. \ref{uso-tfrut} are suggestive of the observed behavior. We show the average home range size $\langle S\rangle$ (measured as fraction of plants used) as a function of the randomness parameter $r$. For most situations of interest in the foraging of animals such as \emph{D. gliroides} it is to be expected that this parameter stays near $r=1$, indicating a strong preference of near plants. But since nothing prevents the consideration of less discerning species, we have explored the whole range from $r=0$ (complete randomness) to $r=1$. Figure \ref{uso-tfrut} shows a decaying behavior of the home range as the rule approaches determinism. This, in addition, is strongly affected by the relaxation time, the fructification rate $\tau$. When $\tau$ is large the resource is depleted more effectively, and the animal needs to cover more territory to feed before the maturation of new fruits relaxes the resource to its stationary value. For this reason the curves are higher for the longer values of $\tau$. For these we observe an abrupt shrinking of the home range size as $r\to 1$. Slower relaxation times display a very different behavior, with a faster transition to a confined home range at a much smaller value of $r$. In other words, a fast relaxation time favors a small or confined home range even for a very non-discriminating behavior in terms of the step size of the walk.

In the case of a lognormal substrate, as expected, there is a confinement of  the movement of the animal to a smaller range within its naturally occurring patches. The transition to confined walks occurs at a small value of $r$ for all relaxation times, even for very slow ones.

\section{A population of non-interacting animals}

If more than one animal share the same substrate, a possible mathematical formulation of the dynamics is of the kind:
\begin{eqnarray}
\frac{du_i}{dt} &=& \mathcal F(u_i,v), \\
\frac{dv}{dt} &=& \mathcal G(\mathbf{u}),
\label{dvdt}
\end{eqnarray}
where $\mathbf{u}=(u_1,\dots)$ is the distribution of the animals and $v$ that of the resource. Both densities evolve in time according to appropriate evolution operators $\mathcal F$ and $\mathcal G$. By solving formally Eq.~(\ref{dvdt}) as $v(t)=v(t,\mathbf{u}(t))$ one can reduce the system to an effectively interacting dynamics within the animal populations, even in the absence of an explicit coupling between the equations for the $u_i$'s:
\begin{equation}
\frac{du_i}{dt} = \mathcal F(u_i, v(t,\mathbf{u}(t))),
\end{equation}
a situation usually called exploitation competition \cite{alatalo87}. The situation is similar to a multispecies dynamics with a shared common resource, even in a mean-field formulation.

In our present study the walks are self and mutually repulsive through the interaction with the substrate, since an animal avoids the plants where the resource has already been depleted. As a consequence, while not directly interacting, the animals feel the presence of one another through the \emph{interference} mediated by the common resource. This produces  both a repulsion and a growth of the home ranges. Two questions are of interest as a first step to analyze the use of a shared space under this circumstances. Firstly, how big does a system need to be to accommodate a population with minimum overlap? Besides, does a little randomness in the step rule (which gives animals the ability to move further) help them in keeping their home ranges apart?

To analyze the interference in a population we make use of the distribution of space usage of each animal. This is defined as the normalized frequency of visits to every plant in the system. By definition, this is a vector in the multidimensional space $\mathbb{R}^N$:
\begin{equation}
f \in S_N \subset \mathbb{R}^N,
\end{equation}
that is, $\sum_k^N (f)_k = 1$, which places $f$ in the simplex $S_N$. The plots shown in Figs.~\ref{tf500-2} and~\ref{aperiodic}, for example, are representations of these vectors, with the size of the blobs proportional to the components of $f$. The home ranges of two animals, then, are characterized by two of these vectors, $f_i$ and $f_j$ (note that these subindices denote animals, while the subindices accompanied with parentheses indicate components in $\mathbb{R}^N$).

A good measure of the overlap of two home ranges is provided by the standard scalar product in $\mathbb{R}^N$:
\begin{equation}
O(f_i,f_j) = \sum_{k=1}^N (f_i)_k (f_j)_k.
\end{equation}
Besides all the good properties of a scalar product the overlap $O$ has the following property, which is of particular interest in the present context: two vectors with the same support (the same plants visited) can have different scalar product. In other words: two home ranges with the same plant support can have different overlaps. This is better understood with a simple example. Suppose that we have two animals on a substrate consisting of three plants. So $f_1=(x_1,x_2,x_3)$ and $f_2=(y_1,y_2,y_3)$. If the two animals share the same plant, for example plant \#1:
\begin{align}
&f_1 = (1,0,0), \\
&f_2 = (1,0,0), \\
\text{then: } &O(f_1,f_2) = 1,
\end{align}
which is the maximum possible overlap. But if the animals share \emph{two} equally frequent plants (say plants \#1 and \#2), the result is different:
\begin{align}
&f_1 = (0.5,0.5,0), \\
&f_2 = (0.5,0.5,0), \\
\text{then: } &O(f_1,f_2) = 0.5,
\end{align}
which is smaller than 1, even though the ``footprint'' of the two animals on the substrate is the same. The usual interpretation for this is a dynamical one, rather than a geometric one. Since they have two plants to browse, they \emph{can} take turns between the plants, thus reducing the interference with respect to the case where they share \emph{just one} plant. For this reason we will refer to $O$ as the \emph{dynamical overlap} below.\footnote{The interpretation of taking turns to share the two plants is not the only possible one. Observe that we say that they \emph{can} take turns. Other dynamical possibilities exist for the same set of $f_i$ and $O$. Still we prefer to refer to $O$ as a dynamical overlap to distinguish it from the one defined below.}

To complement this feature, and to be able to discern when the two home ranges are effectively disjoint (that is, whether the supports of $f_1$ and $f_2$ are disjoint: $\mbox{sup}(f_1) \cap \mbox{sup}(f_2) = \emptyset$), we can use another measure of the overlap, namely the same scalar product divided by the Euclidean norms of $f_i$ and $f_j$:
\begin{equation}
P(f_i,f_j) = \frac{\sum_{k=1}^N (f_i)_k (f_j)_k}{\|f_i\|\,\|f_j\|}.
\end{equation}
In the usual geometric interpretation of the scalar product, $P(f_i,f_j)$ is the cosine of the angle between $f_i$ and $f_j$. This measure of the overlap gives the same value ($P=1$) whenever the footprints of the home ranges coincide. As such, smaller values of $P$ measure an effective spatial separation of the home ranges. Both measures are complementary, as can be seen from the discussion above, so we chose to analyze both, referring to $P$ as the \emph{geometrical} overlap.

\begin{figure}[t]
\begin{center}
\includegraphics[width=\columnwidth]{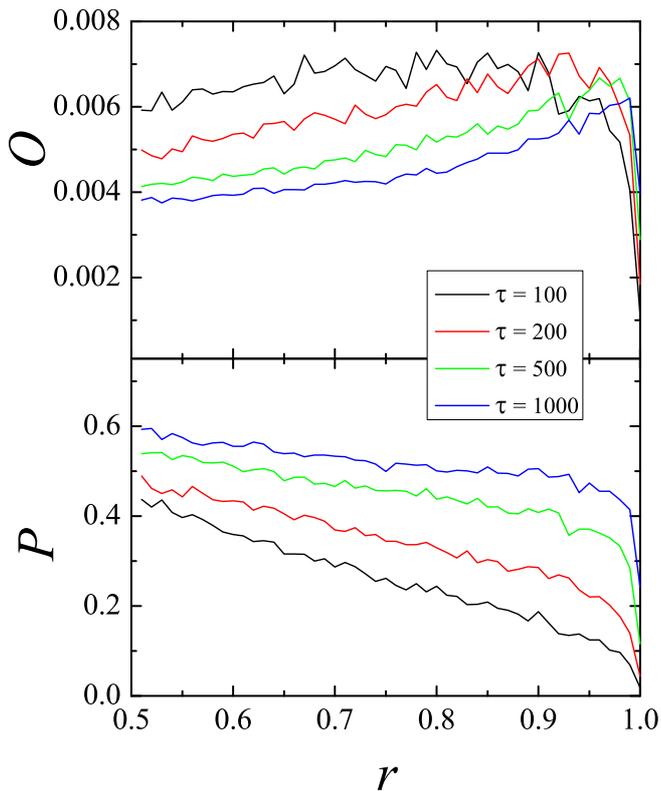}
\end{center}
\caption{Overlaps of the home ranges of two animals as a function of $r$. $N=400$ (uniform substrate). Runs of $10N$ time steps, average of 1000 realizations per value of $r$ (each with a different substrate).} \label{overlaps}
\end{figure}

To begin the analysis of these matters we have studied systems composed of the minimal non-trivial population: two animals. Figure \ref{overlaps} shows the overlaps $O$ and $P$ measured for a range of the probability of choosing the nearest plant, $r$, and for a set of values of the relaxation time $\tau$. Each curve is the result of the average of 1000 realizations per value of $r$, with randomly chosen uniform substrates and initial conditions. The top panel shows $O(f_1,f_2)$ and the bottom one $P(f_1,f_2)$.

Observe that the effect of randomness in the step rule is not the same for the two overlaps. It stands out that a little chance ($r\lesssim 1$) increases the overlaps (because the two home ranges expand due to the smaller $r$). But progressively smaller values of $r$ have different effects on $O$ and $P$. Observe first the dynamical overlap $O$: there is a maximum at an intermediate value of $r$, and a further increase in randomness actually \emph{reduces} the overlap. It can even become \emph{smaller} than the value it has for the deterministic rule, $r=1$.  The same behavior has been observed for all values of $\tau$, but is more striking the slower the ripening, in the sense that a very little randomness in the choice of the step (which is to be expected in most animals) puts the system in the regime of decreasing dynamical overlap.

While this happens to $O$, observe that the purely geometric overlap $P$ grows monotonically when reducing $r$. In other words, the two animals are sharing a common space (high $P$) but taking turns in their use (low $O$). This is a rather surprising behavior in such a simple model. Moreover, it show that there is good reason to keep both definitions of the overlap as complementary descriptions of the use of space.

\begin{figure}[t]
\begin{center}
\includegraphics[width=\columnwidth]{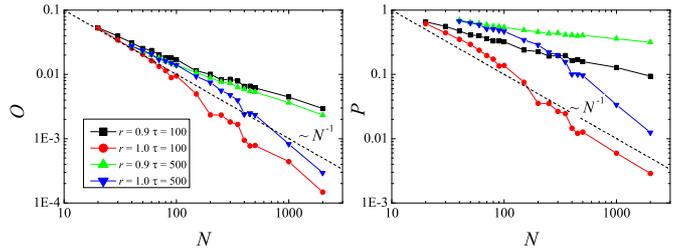}
\end{center}
\caption{Overlaps of two animals as a function of the substrate size, $N$. Runs of $10N$ time steps, average of 100 realizations per value of $r$ (each with a different substrate). Parameters $r$ and $\tau$ as shown in the legend.}
\label{overlaps-np}
\end{figure}

What can be said regarding the first question posed at the beginning of this section, regarding the interference of animals occupying progressively bigger areas? We have analyzed the dependence of the overlaps of the home ranges of two animals in a wide range of systems sizes (defined by the number of plants $N$). The results, corresponding to a set of values of the main parameters of the model, $r$ and $\tau$, are shown in Fig.~\ref{overlaps-np}. The left panel displays the dynamical overlap $P$, while the right one shows the geometric one $O$. Observe firstly that, as expected, in all cases a bigger system accommodates better than a smaller one our minimal population of two animals.

We also observe that both measures of the overlap decay almost algebraically, as a power law of $N$. In the graphics we have added a dashed line that decays as $N^{-1}$. A simple calculation shows that this is the law expected if the home ranges where placed at random without correlation. It is slower than an exponential one for large systems, but shows a faster drop for rather small systems. Our results show that the deterministic walks ($r=1$) follow the $N^{-1}$ behavior for a wide range of system sizes. The departure obeys mainly to a saturation effect in small systems, and to subsampling of the basins of attraction in the bigger ones. It is clear, though, that more realistic walks (slightly random with $r=0.9$) have a very different behavior. They also depend algebraically on the system size, but with a slower decay. For these animals the interference in their home ranges is stronger, compared to non-interacting animals. In other words, two animals interacting in the manner modelled here need a much bigger system than what could be expected from the random overlap of their home ranges.

It is also apparent in our results, and completely reasonable, that a model with $r=1$ (deterministic step) shows smaller overlaps than one with some randomness ($r=0.9$ shown). We see that the former has a drastically smaller overlap for system sizes $N>100$. On the other hand, the dependence on the relaxation rate $\tau$ is less obvious. Compared to the system with $\tau=100$, the plants with $\tau=500$ replenish the resource so slowly that the animals need to browse a bigger fraction of the system. Indeed, the geometric overlap ($O$, right panel) of the slow relaxation system is much greater than the fast one for both the deterministic and the random steps.
This indicates that the \emph{footprints} of their home ranges overlap. Nevertheless, for the same systems the dynamical overlap remains small, of the same order of magnitude that the corresponding systems with the faster $\tau=100$. This shows that, even though the animals need to share a territory because of the depletion of the resource, they can do it with little interference by taking turns in different parts of the corresponding home ranges.

\section{Discussion}

We have analyzed a simple model of animal foraging in an heterogeneous habitat. The movement rules have been kept intentionally simple in order to understand the consequences, in the space use, of a minimal model of foragers exploiting renewable and patchy resources. In this spirit, our model incorporates one essential ingredient in the animal movement (the rule of proximity, complemented with a weaker one of abundance), and one in the resource dynamics (the exponential relaxation).

The properties of the walk defined according to these rules are complex enough for a non-intuitive result. Even if they provide no means for the animal to \emph{remember} the good places to feed, and that there is no \emph{explicit} cost in moving, the combination of closeness and fruitload is enough to guarantee the emergence of a finite home range. The reason for this is that memory of the usage of the landscape is kept in the landscape itself. The walk rule ensures that the animal tends to go away from the current location: it is consuming the resources, so it searches nearby plants for fruit. Yet, the relaxation of the fruitload ensures that the resource is eventually replenished. This is at variance with models of destructive foraging such as \cite{boyer09}, for example, where the visited sites are removed from the system and never revisited, which display non-localized trajectories similar to L\'evy flights. When this happens, the proximity rule (that previously allowed the animal to move away) allows it to come back to places where it has fed before. The peculiarities of the random distribution of patches  and the initial location of the walker determine the sequence in which the plants are visited.

It is particularly notable that, even when the proximity rule is weakened with the inclusion of noise, (the $r<1$ case), there is still a home range, with a well defined probability of occupation of space. This fact stands out as a realistic aspect of our results. It is known that a similar model displays slow (glassy) dynamics, with diverging residence times in cyclic attactors, when randomness is kept below a transition level~\cite{risau03}. The relationship between the two models will be explored in the future.

We have also observed that this phenomenology is robust with respect to details in the distribution of the step choice, even though the numerical results may vary. For example, we have considered two cases (not shown here) that can be thought of as extreme situations of animals with a limited perception of the availability of fruit in their neighborhood. On the one hand, the animal may have an unlimited range, detect all plants with a fruitload under the threshold, and just ignore them in its exponential choice of the one to jump to. On the other hand, if the animal has a very short range of perception of the fruitload, it would choose unsuitable plants (because of low fruitload), discard them, and choose again. The difference is very subtle, but it affects the relative probabilities of plants in the tail of the distribution, and has the effect of slightly inflating the home ranges of the animals that perceive farther. Since the details of these complex behaviors in real animals is difficult to assess completely for the purpose of modelling, it is important that the observed phenomena are the same.

Finally, we have studied the interference of two animals sharing the same resource landscape. This analysis of the common dynamics of two animals is a necessary first step towards the modelling of a larger population. The effective interaction produced by the depletion of the common resource has an effect of inflation of the home ranges, without destroying them. We have characterized this interference by two measures of the overlap of the home ranges, one more dynamical and the other one more geometric. In this context we obtained two main results.

In the first place, the dynamical overlap is non-monotonic with respect to the randomness in the step rule. It has a maximum at a value of $r<1$. In a sense, this corresponds to a better utilization of the resource, through sharing. The evolutionary effects of this pattern cannot be assessed at the present stage of the model, but it certainly will be one of our interests in future developments. In particular, it will be of interest to complement this resource-mediated interference with other known dynamical mechanisms of the origin of territoriality, such as scent deposition~\cite{giuggioli11}.

In the second place, the overlap displays a simple dependence on the system size, namely a power law with a smaller exponent than the one corresponding to non-interacting animals. The other relevant parameters ($r$ and $\tau$) appear involved in this behavior in a no trivial way. In any case, this result points also in the direction of the previous one: two animals are more likely to share the resource (if there is some randomness in the movement rule) than what could be expected from the random overlap of their ranges.

The emergent properties of our simple approach provides a baseline for more realistic models of animals foraging on patchy and renewable resources. An instance where the interaction between animal movement and a dynamical resource plays an important role is the mutualism between plants and their seed dispersers. In the Patagonian temperate forest, for example, there exists a particularly interesting example: the quintral (\textit{Tristerix corymbosus}) and the monito del monte (\textit{Dromiciops gliroides}). \textit{T. corymbosus}, an hemiparasite, depends on agents to disperse their seeds to the branches of potential hosts. It is a keystone species of the temperate forests of southern South America, because  during the winter is the only resource for the hummingbird \textit{Sephanoides sephanoides}, which is one of the most important pollinators in this ecosystem \cite{aiz03}. Furthermore,  the fruits of quintral represent an important food source for the \emph{D. gliroides}, a marsupial endemic to the region and the only current representative of the \textit{Microbiotheria} order. In turn, \emph{D. gliroides} is the only seeds disperser for \textit{T. corymbosus} \cite{amic00}. The study of the relevance of the present findings in such systems is currently under way and will be reported elsewhere.

\begin{acknowledgement}

This work received support from the Consejo Nacional de Investigaciones Cient\'{\i}ficas y T\'ecnicas (PIP 112-200801-00076), Universidad Nacional de Cuyo (06/C304), and Agencia Nacional de Promoci\'on Cient\'{\i}fica y T\'ecnica (PICT-2011-0790).
\end{acknowledgement}

\end{document}